\documentclass[english,10pt,final,twocolumn]{elsarticle}
\usepackage[T1]{fontenc}
\usepackage[latin9]{inputenc}
\usepackage{pdfpages}
\usepackage{amstext}
\usepackage{graphicx}
\usepackage{gensymb}
\usepackage{amssymb}
\usepackage{placeins}
\usepackage{subfigure}
\usepackage[title]{appendix}
\makeatletter
\usepackage{float}
\usepackage{multirow}

\usepackage{blindtext}
\usepackage{titlesec}
\usepackage{setspace}

\titleformat{\section}
  {\bfseries\large}{\thesection}{1em}{}
\titleformat{\subsection}
  {\bfseries}{\thesubsection}{1em}{}
  \titleformat{\appendix}
  {\bfseries\large}{\thesection}{1em}{}

\newcommand{\lyxmathsym}[1]{\ifmmode\begingroup\def\b@ld{bold}
  \text{\ifx\math@version\b@ld\bfseries\fi#1}\endgroup\else#1\fi}
\usepackage{braket}
\usepackage{mathrsfs}
\usepackage{slashed}
\usepackage{bm}
\usepackage{datetime}
\usepackage{siunitx}

\usepackage{amsmath}

\DeclareSIUnit[number-unit-product = {}]\clight{c}
\DeclareSIUnit\eVperc{\eV\per\clight}
\DeclareSIUnit\GeVpercs{\giga\eV\squared\per\clight\squared}
\DeclareSIUnit\MeVpercs{\mega\eV\per\clight\squared}
\journal{Physics Letters B}
\biboptions{numbers,sort&compress}
\usepackage[left=1.6cm,right=1.04cm,top=1.65cm,bottom=1.65cm,columnsep=25pt]{geometry}
\usepackage{lineno}
\usepackage[normalem]{ulem}

\usepackage{hyperref}  
\hypersetup{breaklinks = true, colorlinks = true, allcolors = magenta}
\usepackage{setspace}
\usepackage{graphicx}
\usepackage{wrapfig}

\usepackage{xcolor, soul}
\sethlcolor{yellow}

\makeatother

\usepackage{babel}
\begin{document}

\begin{frontmatter}{}

\title{
Measurement of the helicity-dependent response\\
in quasi-elastic proton knockout from  $^{40}{\rm Ca}$
}

\author[TAU,JSIa]{T.~Kolar\corref{cor2}}
\ead{tkolar@mail.tau.ac.il}
\author[TAU,soreq]{I.~Sabo-Napadensky}
\author[Mainz,JLab]{P.~Achenbach}
\author[Mainz]{M.~Christmann}
\author[Mainz]{M.O.~Distler}
\author[Mainz]{L.~Doria}
\author[Mainz]{P.~Eckert}
\author[Mainz]{A.~Esser}
\author[Pavia]{C.~Giusti}
\author[Mainz]{J.~Geimer}
\author[Mainz]{P.~G\"ulker}
\author[Mainz]{M.~Hoek}
\author[Mainz]{P.~Klag}
\author[TAU]{J.~Lichtenstadt}
\author[Mainz]{M.~Littich}
\author[Mainz]{T.~Manoussos}
\author[Mainz]{D.~Markus}
\author[Mainz]{H.~Merkel}
\author[UL,JSI]{M.~Mihovilovi\v{c} }
\author[Mainz]{J.~M\"uller}
\author[Mainz]{U.~M\"uller}
\author[Mainz]{J.~P\"{a}tschke}
\author[UCR]{S.J.~Paul}
\author[TAU]{E.~Piasetzky}
\author[Mainz]{S.~Plura}
\author[Mainz]{J.~Pochodzalla}
\author[JSI]{M.~Po\v{z}un}
\author[huji]{G.~Ron}
\author[Mainz]{B.S.~Schlimme}
\author[Mainz]{C.~Sfienti}
\author[Mainz]{S.~Stengel}
\author[USK]{E.~Stephan}
\author[USC]{S.~Strauch}
\author[Mainz]{C.~Szyszka}
\author[UL,JSI]{S.~\v{S}irca}
\author[Mainz]{M.~Thiel}
\author[USK]{A.~Wilczek}
\author{\\\textbf{(A1 Collaboration)}}

\cortext[cor2]{Corresponding author}
\fntext[JSIa]{Present address: Jo\v{z}ef Stefan Institute, 1000 Ljubljana, Slovenia.}
\fntext[JLab]{Present address: Thomas Jefferson National Accelerator Facility, Newport News, VA 23606, USA.}

\address[TAU]{School of Physics and Astronomy, Tel Aviv University, Tel Aviv 69978,
Israel.}
\address[soreq]{Soreq NRC, Yavne 81800, Israel.}
\address[Mainz]{Institut f\"ur Kernphysik, Johannes Gutenberg-Universit\"at, 55099
Mainz, Germany.}
\address[Pavia]{INFN, Sezione di Pavia, via A.~Bassi 6, I-27100 Pavia, Italy.}
\address[UL]{Faculty of Mathematics and Physics, University of Ljubljana, 1000
Ljubljana, Slovenia.}
\address[JSI]{Jo\v{z}ef Stefan Institute, 1000 Ljubljana, Slovenia.}
\address[UCR]{Department of Physics and Astronomy, University of California, Riverside, CA 92521, USA.}
\address[huji]{Racah Institute of Physics, Hebrew University of Jerusalem, Jerusalem
91904, Israel.}
\address[USK]{Institute of Physics, University of Silesia in Katowice, 41-500 Chorz\'ow, Poland.}
\address[USC]{University of South Carolina, Columbia, South Carolina 29208, USA.}

\begin{abstract}
The role of the electron-helicity-dependent cross-section term and the structure function $f^{\prime}_{01}$ in the quasi-elastic $A(\vec{e}, e^{\prime}p)$ process was studied. The $f^{\prime}_{01}$ was measured for proton knockout from the $1\mathrm{d}_{3/2}$ shell in $^{40}\mathrm{Ca}$ via the $^{40}{\rm Ca}(\vec{e},e' p)^{39}{\rm K}_{\rm g.s.}$ reaction, leaving the residual nucleus in a well-defined state. It requires a longitudinally polarized electron beam and out-of-plane proton detection.
This structure function vanishes in the absence of final-state interactions (FSI) involving the ejected proton. Presented are the dependencies of $f^{\prime}_{01}$ on the missing momentum (closely related to the initial proton's Fermi momentum) and the angle between the knocked-out proton and the virtual photon momenta. The role of the spin-orbit interaction in FSI through the $\vec{L}\cdot \vec{S}$ term in a nuclear optical potential is discussed.
\end{abstract}
\date{\today}

\end{frontmatter}{}

\section{Introduction}
Quasi elastic $A(e, e^{\prime}p)$ processes, using unpolarized as well as polarized electrons, have been used for decades to investigate properties of the proton bound in the nucleus~\cite{Giusti:1989ww,Kelly:1996hd}. Before a reliable comparison can be made with a free proton, these studies require a good understanding of the target nucleus structure, of the final-state interaction (FSI) between the outgoing proton and the residual nucleus, and of the nuclear current, including  two-body meson-exchange and isobar currents.

We present here a measurement of the so-called \textit{fifth} structure function obtained for quasi-elastic proton knockout in the $^{40}{\rm Ca}(\vec{e},e' p)^{39}{\rm K}_{\rm g.s.}$ process. The fifth structure function is entirely due, and therefore particularly sensitive, to FSI and its measurement is of
particular interest as it allows us to investigate this important contribution to the $(e, e^{\prime}p)$ cross section.

\begin{figure}[ht!]
\begin{center}
    \includegraphics[width=0.85\columnwidth]{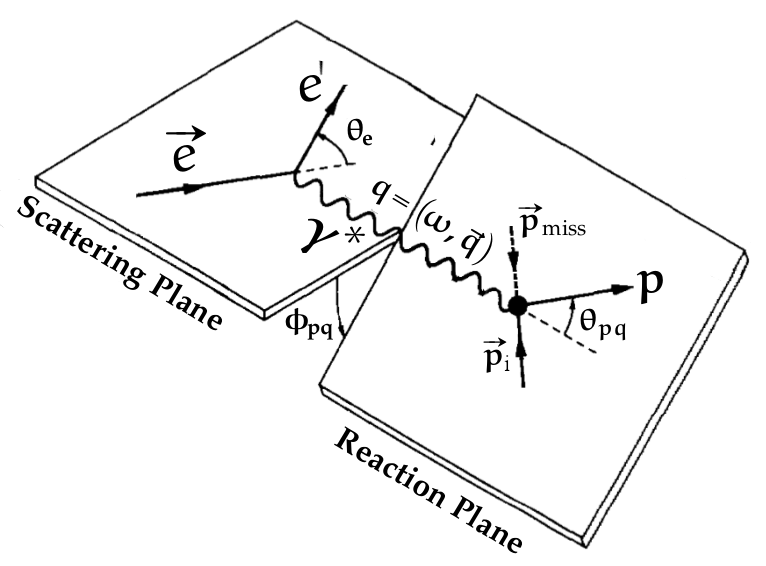}
\end{center}
\caption{
Kinematics of the $(\vec{e},e^{\prime}p\,)$ reaction with the definitions of the kinematic variables.
}
\label{fig:kinematics_diagram}
\end{figure}

The coincidence cross section of the $A(e, e^{\prime}p)$ reaction, where the incoming electron beam is longitudinally polarized with helicity $h$ can be written in the one-photon exchange (Born) approximation as the sum of a helicity-independent ($\Sigma$) and helicity-dependent ($\Delta$) terms as~\cite{Boffi:1996ikg}
\begin{align}\label{eq:CS}
\begin{split}
    \dfrac{d\sigma_h}{d\omega  d\Omega_e d\Omega_p} =& \Sigma+h\Delta \\
    =& K\sigma_{\rm Mott}(\rho_{00}f_{00}+\rho_{11}f_{11}
    +\rho_{01}f_{01}\cos\phi_{pq}\\
    &+\rho_{1-1}f_{1-1}\cos2\phi_{pq}
    +h\rho^{\prime}_{01}f^{\prime}_{01}\sin\phi_{pq})\,,
\end{split}
\end{align}
where the coefficients $\rho_{\lambda\lambda^{\prime}}$ and $\rho^{\prime}_{01}$ depend only on the electron kinematics, $\sigma_{\rm Mott}$ is the Mott cross section, and $K$ combines phase-space and recoil-kinematic factors. The structure functions $f_{\lambda\lambda^{\prime}}$ and the fifth structure function, $f^{\prime}_{01}$, represent the response of the nucleus to the longitudinal and transverse components of the electromagnetic interaction and depend only on the kinematic variables (see Fig.~\ref{fig:kinematics_diagram}) energy and momentum transfer ($\omega$ and $|\vec{q}|$), the outgoing
proton momentum ($p$) and the angle $\theta_{pq}$ between $\vec{q}$ and $\vec{p}$.

If the electron is not polarized only the four functions $f_{\lambda\lambda^{\prime}}$ contribute to the cross section: the longitudinal ($f_{00}$) and transverse ($f_{11}$) and the two interference longitudinal-transverse ($f_{01}$) and transverse-transverse ($f_{1-1}$) structure functions. The fifth structure function, $f^{\prime}_{01}$, contributes only if incoming electron beam is polarized. This function is the imaginary part of the longitudinal-transverse interference between the nuclear charge and the transverse nuclear current; it is related to the antisymmetric part of the hadron tensor and vanishes identically when the hadron tensor is symmetric and real~\cite{Boffi:1996ikg,BOFFI1985697}. The hadron tensor is real whenever the reaction proceeds through a channel with a single dominant phase and at least two interfering complex reaction amplitudes with different phases are necessary to produce a non-vanishing $f^{\prime}_{01}$. In quasi-free nucleon knockout the two interfering channels are proton removal and rescattering through FSI. When FSI is ignored, as in the plane-wave impulse approximation (PWIA), the fifth structure function vanishes identically. As such, a measurement of $f^{\prime}_{01}$ offers a test of the treatment the FSI in theoretical models.

A measurement of $f^{\prime}_{01}$ requires not only a polarized electron beam, but also out-of-plane detection acceptance of the knocked out proton. It results in a cross-section asymmetry between the proton scattering ``above'' ($\phi_{pq}>0$) and ``below'' ($\phi_{pq}<0$) the scattering plane, which is obtained through the measurement of the beam-helicity asymmetry in $(\vec{e},e'p)$ measurements.

Previous measurements of $f^{\prime}_{01}$ were carried out on $^{2}\mathrm{H}$ and $^{12}\mathrm{C}$ at the Bates linear accelerator using the out-of-plane spectrometer (OOPS)~\cite{Mandeville:PhysRevLett.72.3325,Dolfini:1995zz,Dolfini:1999sk}. Those experiments lacked the resolution to separate the different final states of the residual nucleus in the $^{12}\mathrm{C}(\vec{e},e^{\prime}p)^{11}\mathrm{B}$ reaction. However, the fifth structure function depends on the initial and final (residual) nuclear states. In fact, $^{12}\mathrm{C}(\vec{e},e^{\prime}p)$ measurements at the Mainzer Mikrotron (MAMI)~\cite{Kolar:2022} yielded asymmetries of opposite signs for knockout of $1\mathrm{p}_{3/2}$-shell protons with the residual $^{11}\mathrm{B}$ nucleus in $J^{\pi}=(3/2)^-$ or $(1/2)^-$ states~\cite{Kolar:2022}. While $f^{\prime}_{01}$ was not explicitly determined this result implies different $f^{\prime}_{01}$ for each transition.

We present here a measurement of the fifth structure function, obtained for quasi-elastic proton knockout in the $^{40}{\rm Ca}(\vec{e},e' p)^{39}{\rm K}_{\rm g.s.}$ process with a well-defined final state of the residual nucleus ($^{39}\mathrm{K}$ ground state with $J^{\pi}=3/2^+$).
The results are compared with state-of-the-art calculations in the relativistic distorted wave impulse approximation (RDWIA)~\cite{Meucci:2001qc,MeucciPhD}, providing a direct test of the FSI. The comparison also reveals that the spin-orbit part of the nuclear optical potential plays a significant role in determining $f^{\prime}_{01}$. In contrast, it has a much smaller impact on the unpolarized cross section. This suggests that the spin-orbit potential has a major influence on FSI effects in this process.

\section{Experimental setup and kinematics}
The experiment was carried out in 2022 in the A1 experimental hall~\cite{a1aparatus} at MAMI. Spectrometers A and C were used to detect the recoil proton and the scattered electron, respectively. Both spectrometers were set up with their standard detector packages consisting of vertical drift chambers (VDCs), two segmented scintillation layers, and a \v{C}erenkov radiation detector. In spectrometer A the \v{C}erenkov detector was replaced by a focal plane polarimeter (FPP)~\cite{Pospischil:2000pu} used to measure the polarization of knocked-out protons for the purpose of a related analysis~\cite{Kolar:2023}.

The $^{40}{\rm Ca}$ target consisted of three $0.41\,{\rm mm}$ thick foils spaced $15.0\,{\rm mm}$ apart. The three foils were tilted at an angle of $45^{\circ}$ (facing spectrometer A) to minimize the energy loss of the outgoing proton~\cite{Izraeli:2018}. For the $(\vec{e},e^{\prime}p)$ results reported here, we used only events from the central foil to ensure a reliable simulation of spectrometer acceptances.

To determine the polarization of the $600\,{\rm MeV}$ beam, $P_{\rm b}$, and ensure its longitudinal orientation, a M{\o}ller polarimeter~\cite{Wagner} was used. As a cross-check Mott polarimetry~\cite{Tioukine} was performed at the beginning of the experiment and later when the source cathode was replaced. We accounted for variation in quantum efficiency through two distinct periods using two linear fits. The final polarizations ranged from $79.6\,\%$ to $88.0\,\%$.

The data were obtained in near-parallel kinematics ($\vec{p}_{\rm miss}\parallel\vec{q}$) summarized in Table \ref{tab:kinematics}, where the missing momentum is defined as $\vec{p}_{\rm miss}=\vec{q}-\vec{p}$.  In PWIA $\vec{p}_{\rm miss}$ is the opposite of the bound proton's initial momentum. Additional experimental details can be found in Ref.~\cite{Kolar:2023}.

\begin{table}[h!]
\caption{
Central kinematic settings of the $^{40}{\rm Ca}(\vec{e},e^{\prime}p\,)$ reaction measurement of this work.  
}
\begin{center}
\begin{tabular}{lll}
\hline\hline
\multicolumn{3}{c}{Kinematic setting}  \\
\hline
$E_{\rm beam}$ & [MeV]                   &  600 \\
$Q^2$          & [$({\rm GeV}\!/\!c)^2$] & 0.25\\
$p_{\rm miss}$ & [MeV$\!/\!c$]           & $-160$ to $-20$ \\
$p_e$          & [MeV$\!/\!c$]           & 396 \\
$\theta_e$     & [deg]                   & -61.8\\
$p_p$          & [MeV$\!/\!c$]           & 630\\
$\theta_{p}$   & [deg]                   & 40.2\\
\hline\hline
\end{tabular}

\end{center}
\label{tab:kinematics}
\end{table}

\section{Analysis}\label{sec:Analysis}
To ensure the data quality and a good match with acceptance simulation, we applied both quality-of-event-reconstruction and fiducial cuts. The former include an electron-proton coincidence-time cut, $|t_{AC}|<2\,\mathrm{ns}$, a cut on the quality of track reconstruction in the VDCs, and a threshold energy cut in \v{C}erenkov detector for electron identification. Fiducial cuts were applied with respect to the central kinematic settings (Table~\ref{tab:kinematics}) to reduce systematic uncertainties related to the accuracy of the spectrometer optics and acceptance simulation. We cut on the electron and proton momenta, $p^{\rm fiducial}_{e,p}=(1\pm8\%)p_{e,p}$, in-plane angle, $\theta^{\rm fiducial}_{e,p}=\theta_{e,p}\pm3.0^{\circ}$, and out-of-plane angle, $\phi^{\rm fiducial}_{e}=\pm4.0^{\circ}$ and $\phi^{\rm fiducial}_{p}=\pm5.0^{\circ}$. With the spectrometers' target vertex resolutions being $\delta z_{\rm tg,\,A}= 4\,\mathrm{mm}$ and $\delta z_{\rm tg,\,C}= 3\,\mathrm{mm}$ only events with a reaction vertex reconstructed within a $3\delta z$-ellipse around the central foil were considered.  

\begin{figure}[b!]
\includegraphics[width=\columnwidth]{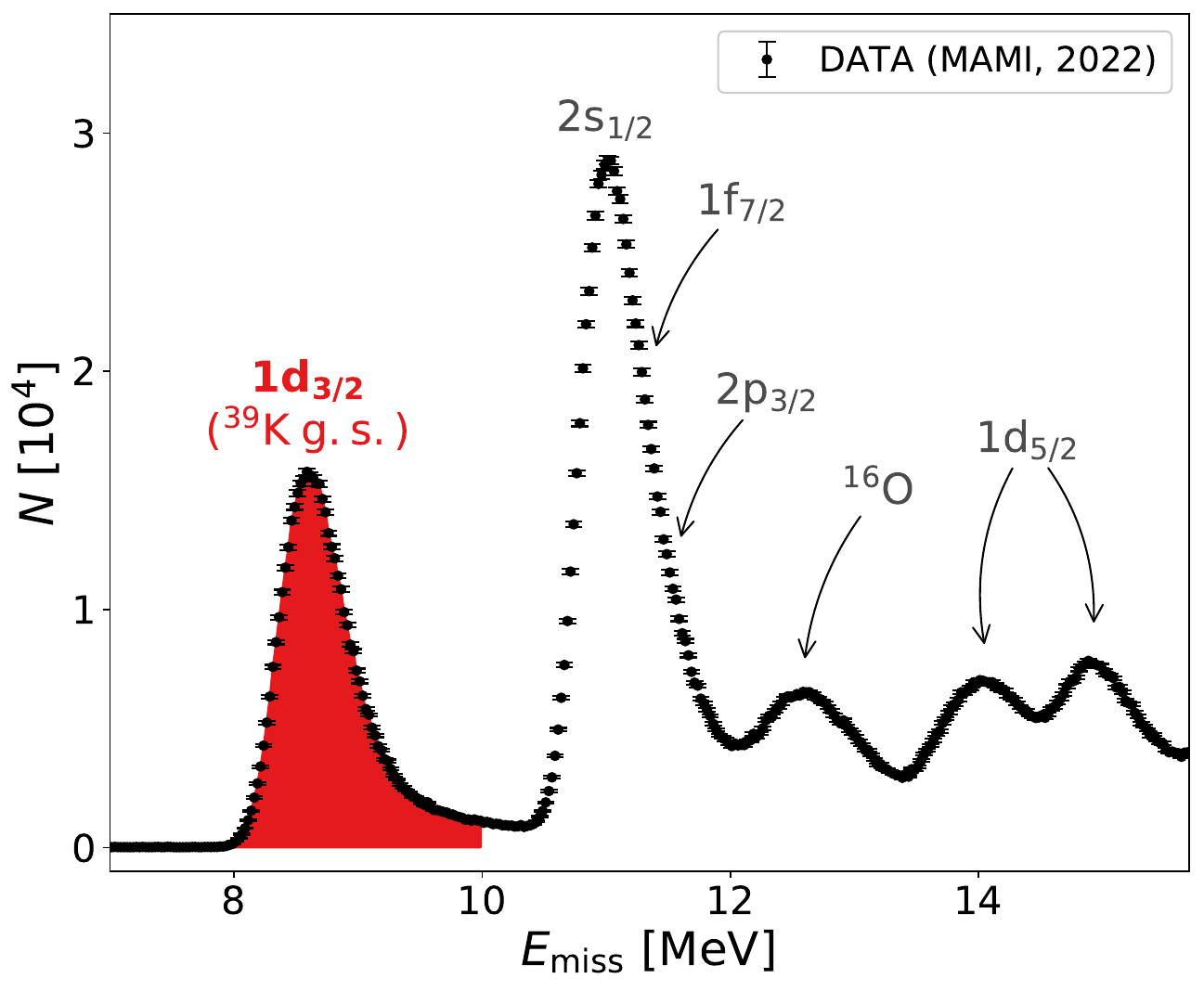}
\caption{
Missing-energy spectrum of the proton knockout from the $^{40}\mathrm{Ca}$ target. Marked is also a peak associated with oxygen contamination due to the target oxidation.
}
\label{fig:emiss_histogram}
\end{figure}

To isolate events with protons originating from the $1\mathrm{d}_{3/2}$ shell of $^{40}\mathrm{Ca}$ and the residual $^{39}\mathrm{K}$ in the ground state, we applied a cut on the missing energy, $8.0\,\mathrm{MeV}\leq E_{\rm miss}\leq10.0\,\mathrm{MeV}$. The missing energy is defined as $E_{\rm miss}\equiv \omega - T_p - T_{{}^{39}{\rm K}}$ with $\omega$ being the energy transfer, $T_p$ the measured kinetic energy of the outgoing proton, and $T_{{}^{39}{\rm K}}$ the calculated kinetic energy of the recoiling $^{39}{\rm K}_{\rm g.s.}$ nucleus. The $E_{\rm miss}$ spectrum is shown in Fig.~\ref{fig:emiss_histogram}. The energy-loss resolution and the radiative corrections are accounted for in the acceptance simulation (see below).

The knockout of a proton from the $2\mathrm{s}_{1/2}$ shell results in a prominent excitation of $J^\pi=\frac{1}{2}^+$ at $2.52\,\mathrm{MeV}$~\cite{KRAMER1989199}. However, it cannot be resolved from the $^{39}{\rm K}$ excited states at $2.82$ and $3.02\,\mathrm{MeV}$ ($\frac{7}{2}^-\text{ and }\frac{3}{2}^-$, respectively). Since the asymmetries of each transition may differ, even having opposite signs~\cite{Kolar:2022}, the $2\mathrm{s}_{1/2}$ state was not analyzed.

The fifth structure function can be determined from the helicity-dependent cross section as
\begin{align}
\begin{split}
    f^{\prime}_{01}&=\dfrac{\Delta}{K\sigma_{\rm Mott}\rho^{\prime}_{01}\sin\phi_{pq}}\\
    &=\frac{\sigma_+ - \sigma_-}{2 P_{\rm b} K\sigma_{\rm Mott} \rho^{\prime}_{01}\sin\phi_{pq}}\\
    &=\frac{N_+/(\mathcal{L}_+\mathcal{A})-N_-/(\mathcal{L}_-\mathcal{A})}{2 P_{\rm b} K\sigma_{\rm Mott} \rho^{\prime}_{01}\sin\phi_{pq}}\,,    
\end{split}
\end{align}
where $N_{+,-}$ is the total number of events attributed to the knock-out of $1\mathrm{d}_{3/2}$ protons for either positive or negative helicity states and $\mathcal{L}_{+,-}$ is the associated integrated luminosity corrected for the coincidence dead time of the trigger. $\mathcal{A}$ is the detector acceptance. 
Since $\mathcal{L}_+= \mathcal{L}_- = \mathcal{L}/2$, we get
\begin{equation}\label{eq:fifth}
    f^{\prime}_{01}=\frac{(N_+ - N_-)/(\mathcal{L}\mathcal{A})}{P_b K\sigma_{\rm Mott} \rho^{\prime}_{01}\sin\phi_{pq}}\,.
\end{equation}
The denominator was calculated on a per-event basis and averaged across a given bin. We followed the convention from Ref.~\cite{MeucciPhD} to determine $K$ and $\rho^{\prime}_{01}$. 

Considering only the knock-out of $1\mathrm{d}_{3/2}$ protons that appear as a peak in the missing energy spectrum, the acceptance integration can be performed in five dimensions. This integration is performed by using \texttt{Simul++} software~\cite{MAMI-DAQ-2001} starting with an isotropic generation of events sampling the volume $\Delta\omega  \Delta\Omega_e \Delta\Omega_p$ and calculation of the outgoing proton momentum, $p_p$, as determined by the discrete missing energy. For each generated event \texttt{Simul++} accounts for target energy losses, radiative corrections, and spectrometers' resolutions. The final acceptance is determined with the standard Monte-Carlo technique by calculating
\begin{equation}
         \mathcal{A} = \dfrac{N_{\rm Accepted}}{N_{\rm Thrown}}\Delta\omega  \Delta\Omega_e \Delta\Omega_p\,. 
\end{equation}

While an explicit determination of the unpolarized experimental cross-section is not needed to obtain the fifth structure function, it is used to extract the experimental spectroscopic factor for an overall adjustment of the RDWIA calculation to the measured data. The same factor is required when the calculations of $f^{\prime}_{01}$ are compared to the experimental data. To evaluate the measured unpolarized cross-section from~(\ref{eq:CS}), we use
\begin{equation}
    \Sigma=\dfrac{N}{\mathcal{L}\mathcal{A}}\,.
\end{equation}
All RDWIA calculations were performed by sampling the kinematic parameters of the isotropically generated events that passed the acceptance simulation, accurately representing the kinematic phase space covered by the data.

\begin{figure}
\includegraphics[width=\columnwidth]{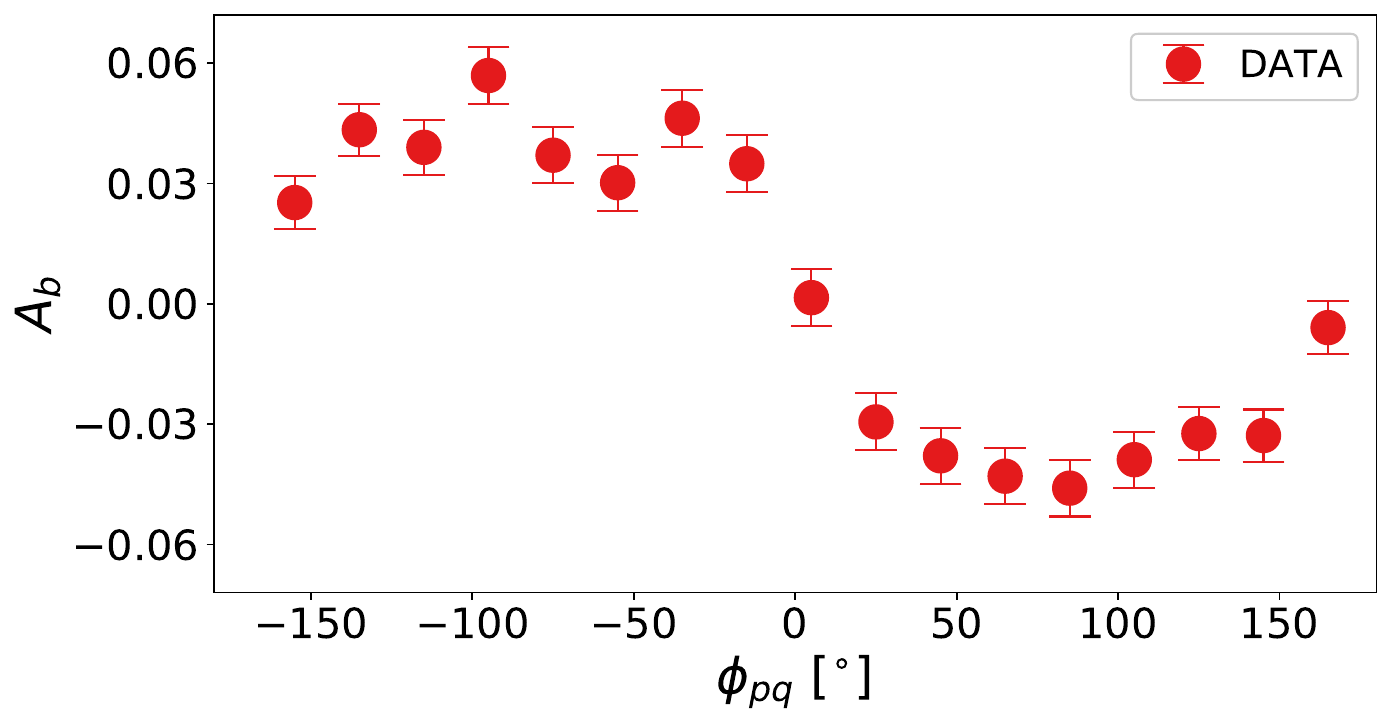}
\caption{The electron-helicity asymmetry from eq.~(\ref{eq:Ab}) for $1\mathrm{d}_{3/2}$ knockout from $^{40}\mathrm{Ca}$ as a function of the out-of-plane angle $\phi_{pq}$.}
\label{fig:asymm_vs_phipq}
\end{figure}

We note that unlike previous measurements that focused on $\phi_{pq}=90^{\circ}$, our acceptance covers the full $\phi_{pq}$ range. The $\phi_{pq}$ dependence of the measured asymmetry is shown in Fig.~\ref{fig:asymm_vs_phipq}. 

Using (\ref{eq:fifth}) we do not rely on asymmetry amplitude determination, 
\begin{align}\label{eq:Ab}
    &A_b=\dfrac{\Delta}{\Sigma}=\dfrac{\sigma_+/2-\sigma_-/2}{\sigma_+-\sigma_-}=\dfrac{N_+ - N_-}{N_+ + N_-}\\
    &=\dfrac{\rho^{\prime}_{01}f^{\prime}_{01}\sin\phi_{pq}}{\rho_{00}f_{00}+\rho_{11}f_{11}
    +\rho_{01}f_{01}\cos\phi_{pq}
    +\rho_{1-1}f_{1-1}\cos
    (2\phi_{pq})}\,, \nonumber
\end{align}
to determine $f^{\prime}_{01}$. Extracting the amplitude of the asymmetry can be challenging because some contributions from other structure functions depend on $\phi_{pq}$ as well; hence, one cannot assume a simple $\sin\phi_{pq}$-dependence.

To investigate possible systematic uncertainties, we also evaluated the asymmetry for elastic scattering off the hydrogen from the residual storage oil on the target to be $0.0011\pm0.0014$. For the cross-section the main source of systematic uncertainty comes from the effective luminosity determination. Target thickness, target angle, and beam current precision are all contributors, but due to very high rates in the spectrometers, the largest contribution comes from the trigger dead time determination. The combined uncertainty of the luminosity is $15.3\,\%$. The efficiency of the scintillation detectors is $(99.0\pm 1.0)\,\%$~\cite{Mihovilovic:2024ymj}. The efficiency of the track reconstruction with the VDCs was found to be $(99.9\pm 0.1)\,\%$. The efficiency corrections apply for each spectrometer separately and were considered a multiplicative correction factor to the data. An additional uncertainty of $2.4\,\%$ comes from the $E_{\rm miss}$ cut due to small inaccuracies in the simulation of the detector resolution and radiative tail. The $f^{\prime}_{01}$ measurement is also affected by the uncertainty in the determination of the beam of $2.0\,\%$. This results in the total systematic uncertainty of $15.6\,\%$ for the cross-section and $15.7\,\%$ for the fifth structure function. We note that the largest contributions due to the luminosity determination and $E_{\rm miss}$ cut impact only the spectroscopic factor uncertainty and not the agreement between the calculation and the data.

\section{Results}
\begin{figure}[t!]
\includegraphics[width=\columnwidth]{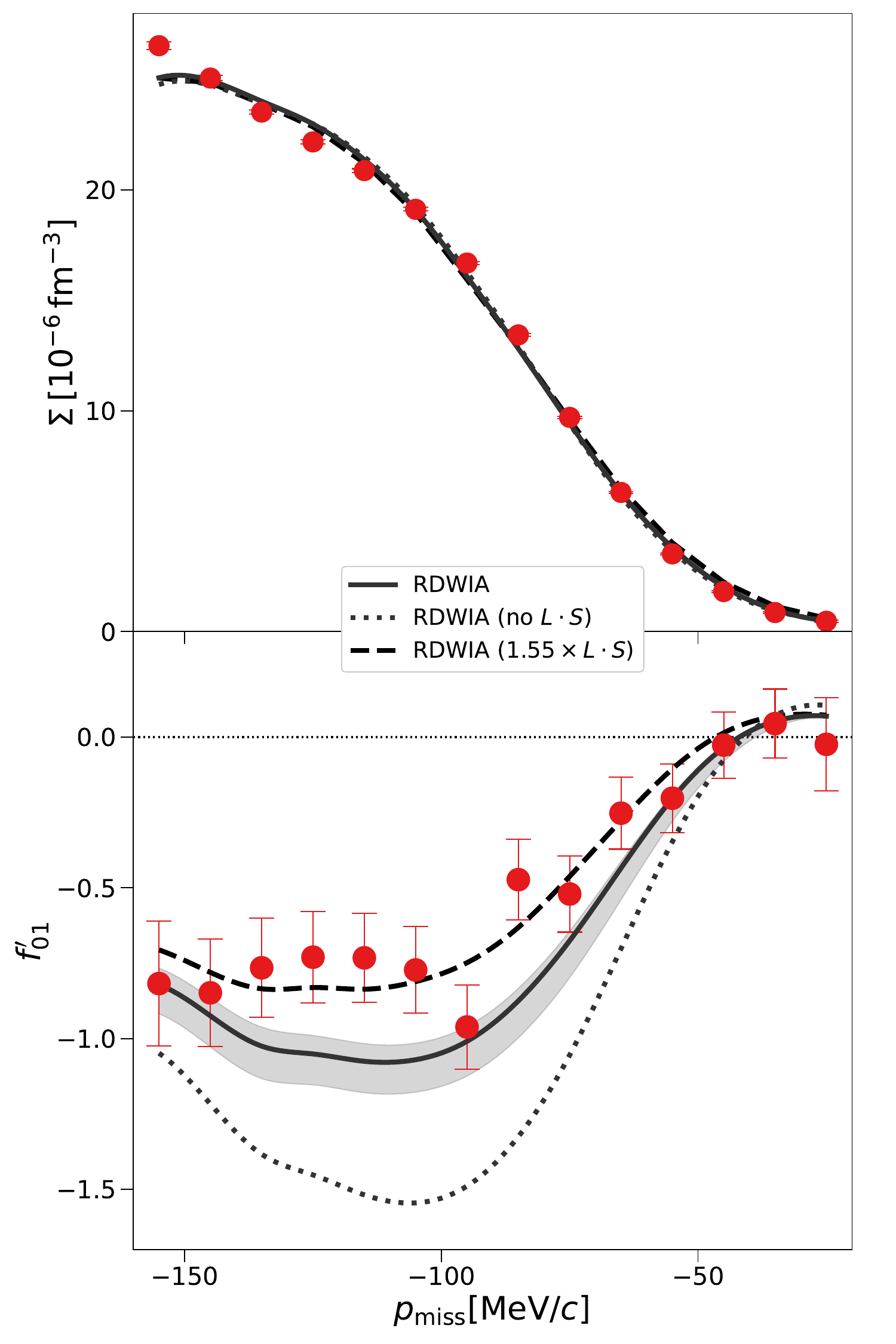}
\caption{Measured unpolarized cross section (top) and fifth structure function (bottom) as a function of missing momentum, $p_{\rm miss}$, with associated statistical uncertainties. Solid line is RDWIA calculation with variation due to use of different ingredients shown as a gray band. Show is also RDWIA result with excluded (dotted) and enhanced (dashed) $\vec{L}\cdot \vec{S}$ term in the nuclear optical potentials. A spectroscopic factor of $0.52$ was applied to all RDWIA calculations except the one with modified $\vec{L}\cdot \vec{S}$ term contribution where specroscopic factor of $0.49$ was used. }
\label{fig:Pmiss}
\end{figure}

\begin{figure}[t!]
\includegraphics[width=\columnwidth]{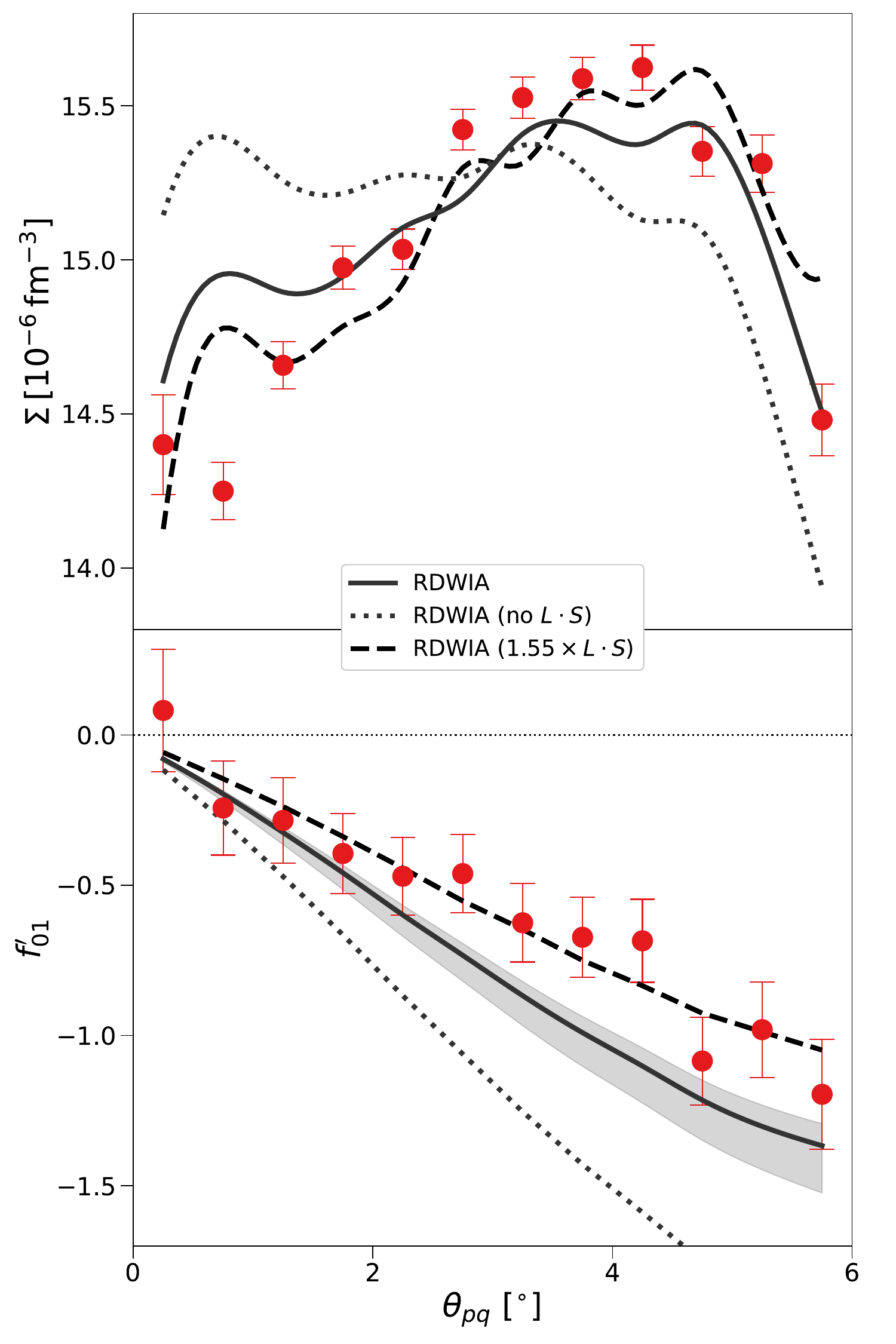}
\caption{Same as Fig.~\ref{fig:Pmiss} but as a function of the polar angle, $\theta_{pq}$, between the outgoing proton and momentum transfer. For RDWIA calculations same spectroscopic factor is used as in Fig.~\ref{fig:Pmiss}. }
\label{fig:Theta}
\end{figure}

To compare the measured and calculated fifth structure function, $f^{\prime}_{01}$, 
we first had to obtain the spectroscopic (reduction) factor from the comparison between measured and RDWIA calculated unpolarized cross
sections, $r_{\rm SF}=\Sigma_{\rm Data}/\Sigma_{\rm Calc}$. Furthermore, a comparison of the measured and calculated $\Sigma$ serves as a test of the quality of acceptance simulation required to evaluate Eq.~(\ref{eq:fifth}). The cross-sections for the $^{40}\mathrm{Ca}(e,e^{\prime}p)^{39}\mathrm{K_{g.s.}}$ reaction as functions of $p_{\rm miss}$ and $\theta_{pq}$ are shown in the top panels of Figs.~\ref{fig:Pmiss} and~\ref{fig:Theta} along with RDWIA calculations (solid lines). 

The RDWIA calculation uses the democratic optical potential~\cite{Cooper:2009}, the bound-state wave function from Ref.~\cite{SHARMA1993377}, the \textit{cc2} prescription for the nuclear current from Ref.~\cite{DEFOREST1983232}, and the free-proton electromagnetic form factors from Ref.~\cite{Bernauer}. These are state-of-the-art calculations that agree well with the results of both unpolarized $(e,e^{\prime}p)$~\cite{Meucci:2001qc} and polarization-transfer measurements~\cite{Izraeli:2018,Brecelj:2020,Kolar:2020,Kolar:2023}.

The calculations are adjusted to the measured cross section with a spectroscopic factor of $0.52$, obtained from the comparison over the $p_{\rm miss}$ dependence. The same factor is applied when comparing the calculation with the measured $\theta_{pq}$ angular dependence. 


The measured $f^{\prime}_{01}$, obtained from (\ref{eq:fifth}), is shown in the bottom panels of Figs.~\ref{fig:Pmiss} and~\ref{fig:Theta}, where its $p_{\rm miss}$ and $\theta_{pq}$ dependencies are compared to RDWIA calculations. The dependence of the fifth structure function on $\theta_{pq}$ is shown using the entire dataset where $-160<p_{\rm miss}<-20\,\mathrm{MeV/c}$. Splitting the data into low- and high-$p_{\rm miss}$ ranges showed no significant difference in the $\theta_{pq}$ dependence. Similarly, the $p_{\rm miss}$ dependence was analyzed over the full range of $\theta_{pq}$ ($0^{\circ}\leq \theta_{pq}\leq6^{\circ}$).

The gray bands in the figures of $f^{\prime}_{01}$ indicate the sensitivity of the RDWIA calculations to different ingredients. We used i) two alternative parameterizations of the relativistic optical potential; Energy-Dependent A-Independent (EDAI) fit to $^{40}\mathrm{Ca}$ data and Energy-Dependent A-Dependent (EDAD) fit~\cite{Cooper:1993nx}, ii) another bound-state wavefunction~\cite{Lalazissis:1996rd}, and iii) different off-shell nuclear current prescriptions (\textit{cc1}, \textit{cc3})~\cite{DEFOREST1983232}. The band was obtained by using the same spectroscopic factor for the different calculations to isolate the sensitivity of the calculated $f^{\prime}_{01}$ to the calculation ingredients. Individual calculations are shown in Fig.~S1 [supplemental material]. The calculations, each adjusted with its spectroscopic factor, for $\Sigma$ and $f^{\prime}_{01}$ are shown in Fig.~S2.

The dependencies of both $\Sigma$ and $f^{\prime}_{01}$ on $p_{\rm miss}$ and $\theta_{pq}$ are in decent agreement between the data and the theoretical RDWIA calculations (solid). However, it seems that the calculations consistently overestimate the strength of $f^{\prime}_{01}$. Using different optical potentials and/or wavefunctions results in relatively small variations (gray band). Also shown is a calculation in which the $\vec{L}\cdot \vec{S}$ term of the potential was switched off (dotted). The $\vec{L}\cdot \vec{S}$ term has a significant impact on $f^{\prime}_{01}$. Its contribution is positive and reduces the generally negative amplitude of $f^{\prime}_{01}$. The unpolarized cross sections are much less sensitive to the spin-orbit interaction, but one can observe its effect on its angular dependence.

The role of the $\vec{L}\cdot \vec{S}$ term in the optical potential and its effect on FSI have been the subject of many studies. It was shown to have a large contribution in induced polarization in the quasi-elastic $(e,e^{\prime}\vec{p})$ process~\cite{Paul:2020}, as well as influence various hadron scattering processes~\cite{LeeSchiffer1964,LICHTENSTADT1978}. Our results suggest a strong sensitivity of $f^{\prime}_{01}$ to the $\vec{L}\cdot \vec{S}$ interaction, consistent with our measurements of the electron-helicity asymmetry on $^{12}\mathrm{C}$ and $^{2}\mathrm{H}$~\cite{Kolar:2022}. 

Although we do not attempt to refit the optical potential parameters, which would require a plethora of data, our results suggest that $f^{\prime}_{01}$ is a very useful tool to study the $\vec{L}\cdot \vec{S}$ term contribution within RDWIA calculations. It seems that the contribution from the $\vec{L}\cdot \vec{S}$ term reduces the $f^{\prime}_{01}$ amplitude and has almost no effect on the $p_{\rm miss}$ dependence of the cross section, but modifies its $\theta_{pq}$ dependence with distinct behavior below and above 4 degrees. We repeated the calculations with an adjusted $\vec{L}\cdot \vec{S}$ term in the optical potential to match the measured $f^{\prime}_{01}$ resulting in an enhancement factor of $1.55$. The results presented in Figs.~\ref{fig:Pmiss} and~\ref{fig:Theta} demonstrate a significantly improved agreement between the calculations and the experimental data for both, the $p_{\rm miss}$ and $\theta_{pq}$ dependencies of $f^{\prime}_{01}$ reducing the $\chi^2/\mathrm{DoF}$ from $2.16$ to $0.49$ and $2.53$ to $0.42$, respectively. Notably, the $p_{\rm miss}$ dependence of the cross-section and the associated spectroscopic factor ($r_{\rm SF}=0.49$) hardly change, while the agreement between the calculated and measured angular distribution improves.
 



\section{Conclusions}
We presented here a first measurement of the fifth structure function, $f^{\prime}_{01}$, in the $(\vec{e},e^{\prime}p)$ process for the knockout of a $1\mathrm{d}_{3/2}$ proton in $^{40}\mathrm{Ca}$, to the residual system of $^{39}\mathrm{K}$ in its ground state ($J^{\pi}=\frac{3}{2}^+$). It is obtained from the helicity difference in the $(\vec{e},e^{\prime}p)$ cross section. The measured $f^{\prime}_{01}$ is in agreement with RDWIA calculations and vanishes in the absence of FSI. It is less sensitive to the details of the nuclear structure~\cite{vdsluyis:94} as is indicated by the gray bands in Figs.~\ref{fig:Pmiss} and~\ref{fig:Theta}. The substantial contribution of the $\vec{L}\cdot \vec{S}$ term to $f^{\prime}_{01}$ suggests that it plays an important role in FSI and the re-scattering dynamics of the proton ejectile. This finding highlights $f^{\prime}_{01}$ as  a valuable tool for probing and refining the impact of the $\vec{L}\cdot \vec{S}$ interaction in hadronic processes. 

\section{Acknowledgements}
We would like to thank the Mainz Microtron operators and technical crew for the excellent operation of the accelerator during the challenging times of worldwide pandemic. This work is supported by the Israel Science Foundation (Grant 951/19) of the Israel Academy of Arts and Sciences, by the Israel Ministry of Science, Technology and Space, by the PAZY Foundation (Grant 294/18), by the the PRISMA+ (Precision Physics, Fundamental Interactions and Structure of Matter) Cluster of Excellence, by the Deutsche Forschungsgemeinschaft (Collaborative Research Center 1044), by the Federal State of Rhineland-Palatinate, by the U.S. National Science Foundation (PHY-2111050), and by the  United States-Israeli Binational Science Foundation (BSF) as part of the joint program with the NSF (grant 2020742). We also acknowledge the financial support from the Slovenian Research Agency (research core funding No.~P1\textendash 0102) and the Research Excellence Initiative of the University of Silesia in Katowice.  
\FloatBarrier
\bibliographystyle{elsarticle-num}

\addcontentsline{toc}{section}{\refname}\small{\bibliography{hdep}}
\clearpage

\end{document}